\documentclass[twocolumn,epj]{webofc}
\usepackage[varg]{txfonts}

\woctitle{International Workshop on Radiopure Scintillators RPSCINT 2013}

\begin{document}

\title{Semi-empirical calculation of quenching factors for scintillators: new results}

\author{V.I. Tretyak \inst{1}\fnsep\thanks{\email{tretyak@kinr.kiev.ua}}} 

\institute{Institute for Nuclear Research, MSP 03680 Kyiv, Ukraine}

\abstract{
New results of calculation of quenching factors for ions in scintillators in semi-empirical
approach described in \cite{Tre10} are presented. In particular, they give additional
arguments in favour of hypothesis that quenching factors for different particles can be described 
with the same Birks factor $kB$, if all the data were collected in the same conditions 
and processed in the same way.
} 

\maketitle

\section{Introduction}

It has long been experimentally known \cite{Bir51} that light yield ($L$) of scintillators for 
incoming heavy particles (protons, $\alpha$ particles, heavier ions) is lower than that for 
electrons of the same energy: $L_i = QF \cdot L_e$, where $QF$ is the quenching factor. 
It depends on many conditions:
(1) scintillating material itself (with specific impurities and defects);
(2) type and quantity of dopants (if any);
(3) temperature; 
(4) type of particle ($p$, $\alpha$, heavier ions);
(5) particle's energy;
(6) electric fields (if any);
(7) conditions of measurements (including e.g. time of collection of scintillation signals because
different particles produce faster or slower scintillation response).
Because of these dependencies, we do not consider $QF$ (and related Birks factor $kB$, see later)
as some fundamental constant of a given
scintillating material (as it is quite often considered) but as a variable associated 
with specific set-up, conditions of measurement and data accumulation and processing.
The $QF$ values are usually in the range from $\simeq 0.8$ for protons to $\simeq0.02$ for heavy ions
(e.g. W). More details and examples can be found in \cite{Tre10}.

Knowledge of quenching factors is very important in sensitive searches for Weakly Interacting 
Massive Particles (WIMPs) \cite{Fre13} through their scattering on atomic nuclei incorporated in 
massive scintillators or scintillating bolometers installed deep underground, like in the DAMA/LIBRA 
dark matter (DM) experiment with 250 kg of NaI(Tl) \cite{Ber13}. 
But, generally speaking, information on $QF$ values is necessary in any measurement of signals from 
ions with the help of scintillators.

To measure quenching factors, ion's beams are used (see e.g. \cite{Biz12}), or monoenergetic neutron
sources (e.g. \cite{Gas12}) which allow to create nuclear recoils with known energies in a 
bulk of a scintillator. Theoretical models were developed in papers \cite{Mur61,Lin63,Hit05,Mei08} 
but any of them does not allow to predict $QF$ for all detectors and for any particle at any energy 
(and very often even to describe already measured experimental data). 

In these circumstances, some semi-empirical approaches, which allow to describe measured data and 
predict the needed $QF$'s, are valuable. One of the methods, built on semi-empirical formula of 
Birks \cite{Bir51}, was presented in \cite{Tre10}. Being quite simple and based on publicly
available software, it nevertheless allowed to successfully describe $QF$'s for many measurements
with scintillators of different kinds. Here we present new results obtained after publication of
Ref. \cite{Tre10}.

\section{Outlines of the method}

Quenching factor for ion is calculated as a ratio of light yield for ion to that for electrons:

\begin{equation}
QF_i(E) =
\frac{L_i(E)}{L_e(E)} =
\frac{\int_0^E \frac{dE}{1+kB(\frac{dE}{dr})_i}}
     {\int_0^E \frac{dE}{1+kB(\frac{dE}{dr})_e}},
\end{equation}

\noindent where $(dE/dr)_i$ and $(dE/dr)_e$ are stopping powers (SP) for ions and electrons, 
respectively, and $kB$ is the so-called Birks factor. 
It is clear that the result depends on approximations used in calculations of $(dE/dr)_i$ and $(dE/dr)_e$.
Stopping powers for ions are calculated here with the SRIM code \cite{SRIM}, 
and for electrons with the ESTAR program \cite{ESTAR}. Both codes are
publicly available, easy to use and, in fact, present one of the best software in the field.
In both cases, total stopping powers are used for $QF$ calculations.

It is possible to obtain the following approximation (see \cite{Tre10}) for the quenching factor:

\begin{equation}
QF_i(E) \simeq \frac{1}{kB(dE/dr)_i}
\end{equation}

\noindent which demonstrates that $QF_i$ depends on energy and is minimal when $(dE/dr)_i$ is 
maximal. 

In the approach presented above, quenching factors depend only on one parameter $kB$ which can
be found from fit of some experimental $QF$ data. This has as some demerits, because bigger
number of parameters usually allow to describe experimental data in a better way, 
but also merits, because in ideal case only one experimental point allows to find the $kB$ value
and after to calculate $QF$ with this $kB$ for all other particles and all energies of interest.

\section{Results}

Below we present some new results obtained after publication of work \cite{Tre10}.

(1) {\it Ar ions in liquid Ar.} 
Quenching factors for Ar ions in liquid Ar were measured 
in the MicroCLEAN studies in \cite{Gas12} in the energy range 11 -- 239 keV (see Fig.~1). 
The data at energies above 20 keV were approximated in \cite{Gas12}
just by constant value $QF=0.25$, and it was written that ``An observed upturn in the scintillation 
efficiency below 20 keVr is currently unexplained.'' Fig.~1 shows also curve calculated with 
Eq.~(1) with the value $kB = 1.4$ mg/(MeV$\cdot$cm$^2$); 
one can see that it describes also the upturn below 20 keV, not ideally but better than just a constant.  

\begin{figure}
\centering
\includegraphics[width=8.1cm]{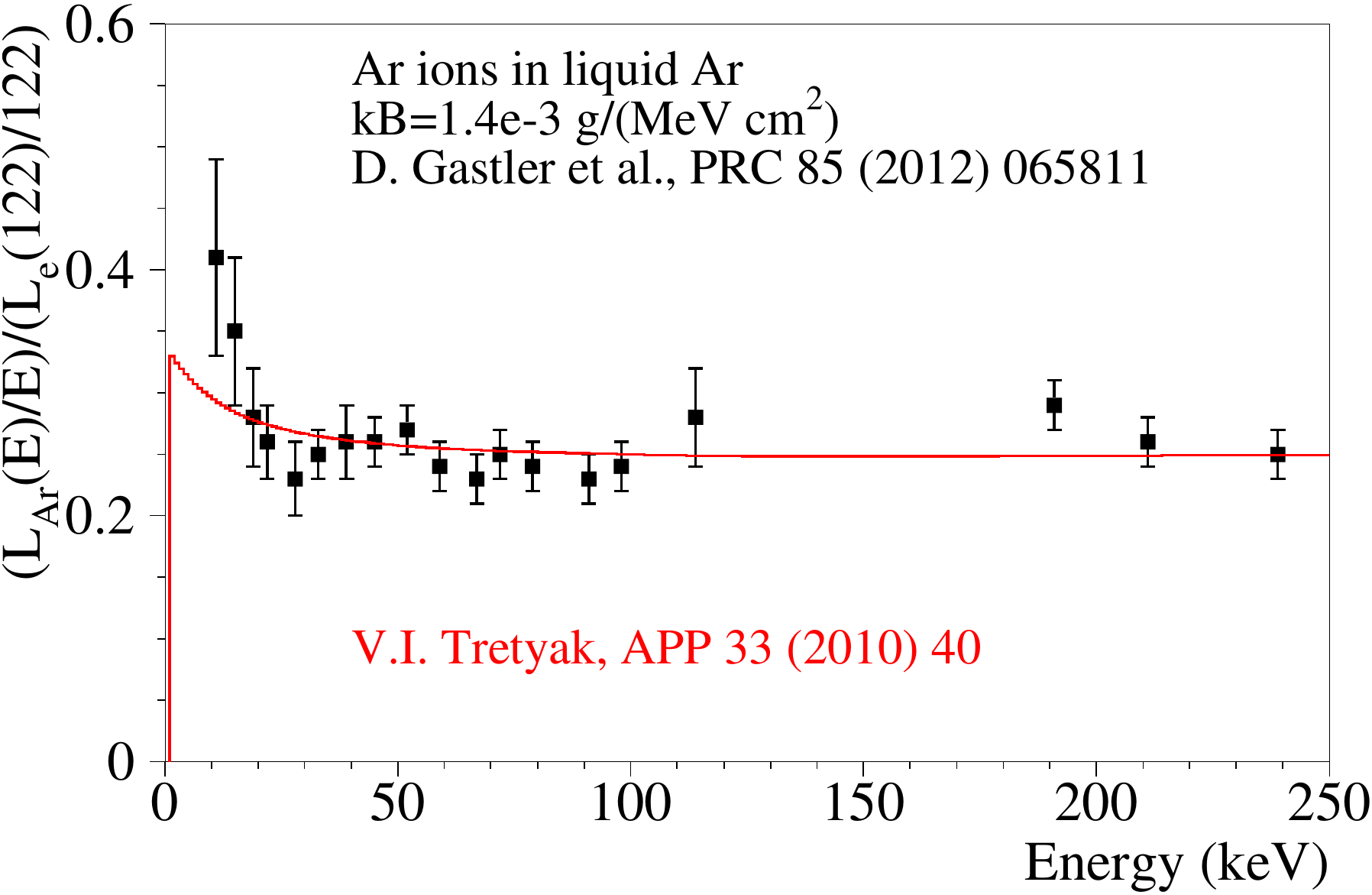}
\caption{Quenching factors for Ar ions in liquid Ar: experimental data \cite{Gas12}
together with calculations with Eq.~(1).}
\end{figure}

Quenching factors for Ar ions in liquid Ar were also measured recently inside the DARWIN framework 
\cite{Reg12}. They are shown in Fig.~2a together with data \cite{Gas12} and two theoretical 
models of Lindhard \cite{Lin63} and Mei \cite{Mei08}. Fig.~2b compares these data with
calculations in accordance with Eq.~(1) with $kB = 1.25$ mg/(MeV$\cdot$cm$^2$).

\begin{figure}
\centering
\includegraphics[width=7.8cm]{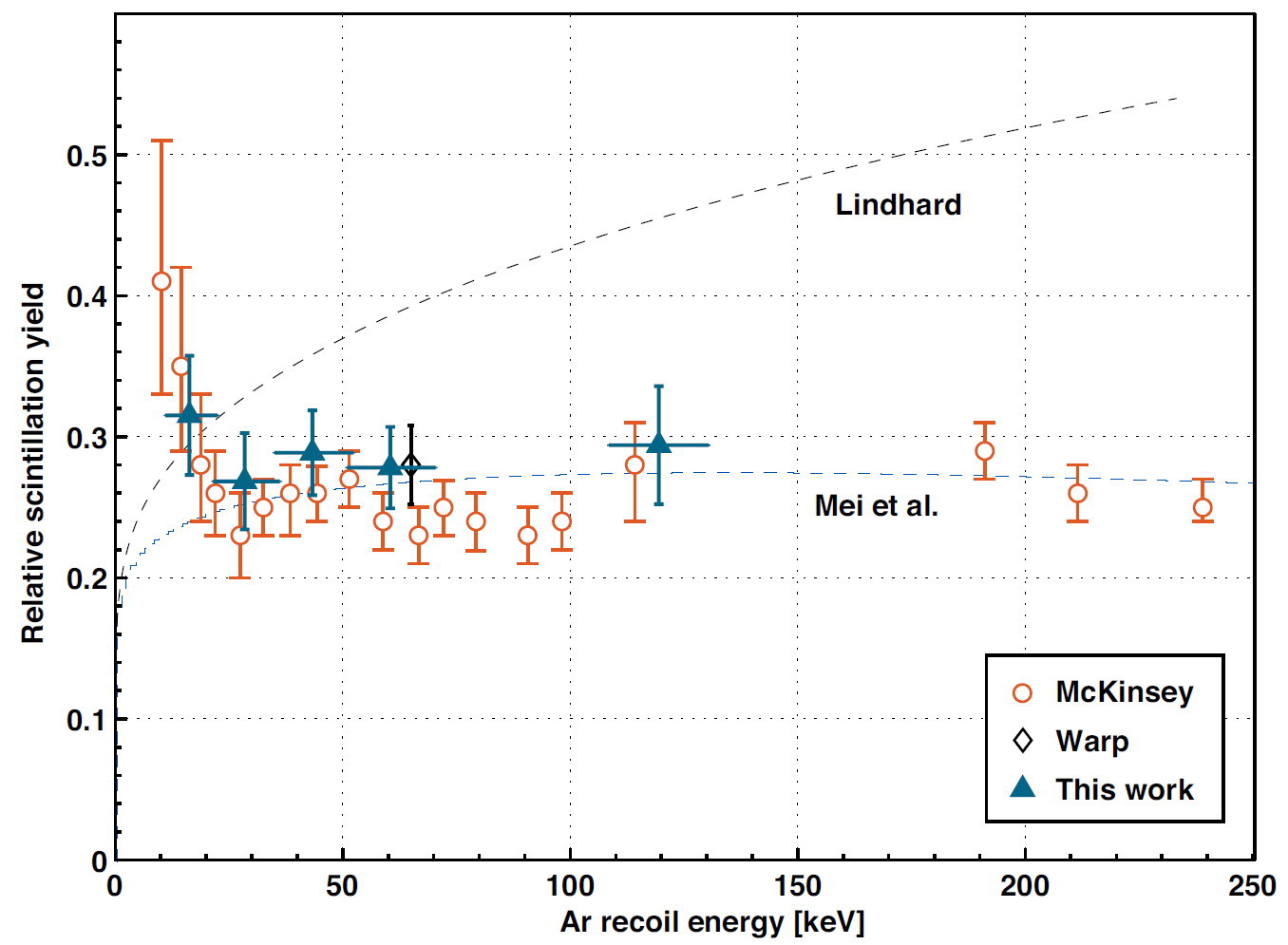} \\
\includegraphics[width=8.1cm]{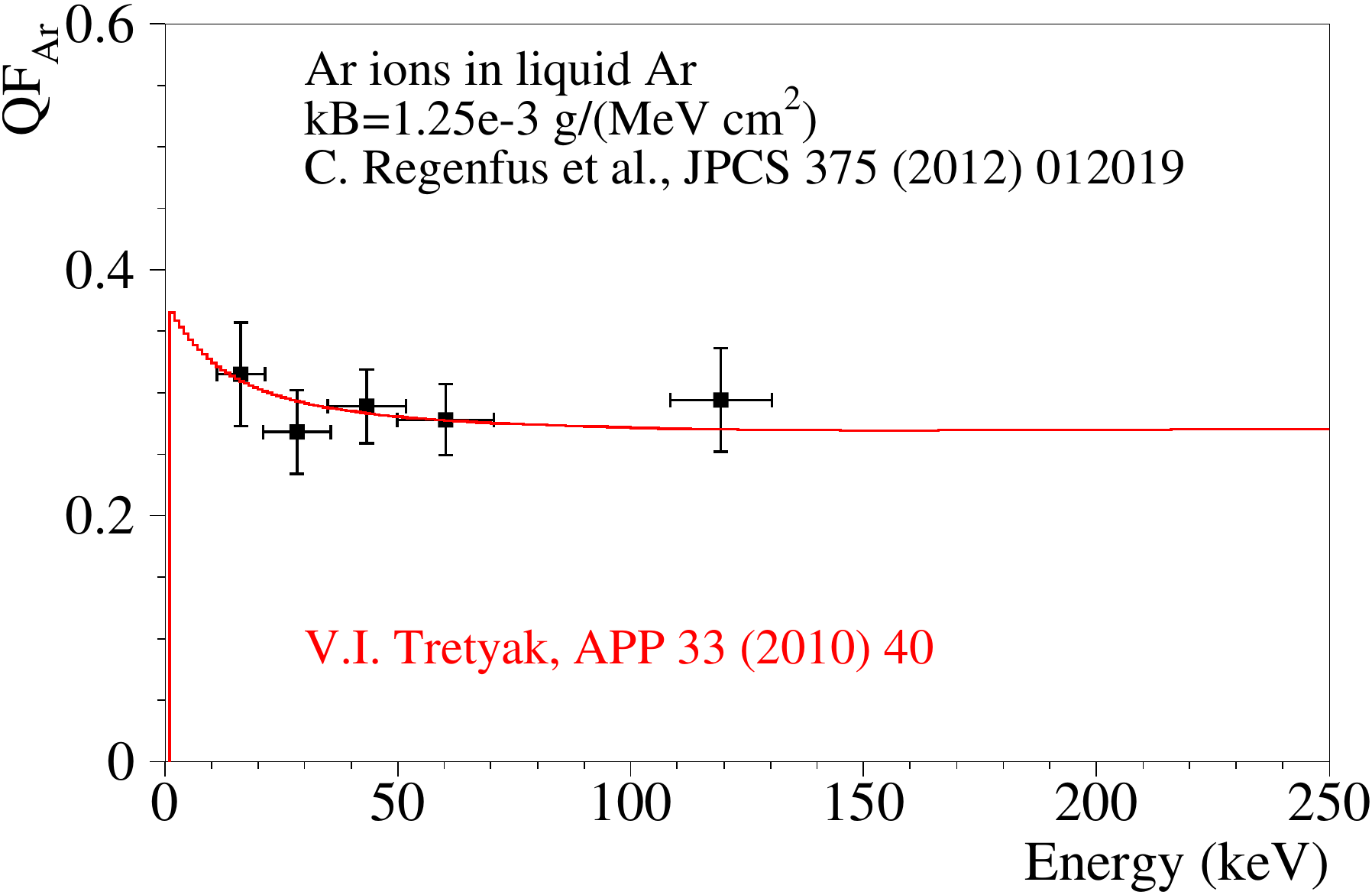}
\caption{Quenching factors for Ar ions in LAr: (top) experimental data \cite{Gas12,Reg12}
together with theoretical curves \cite{Lin63,Mei08} (Fig.~8 of Ref.~\cite{Reg12}); 
(bottom) data \cite{Reg12} together with curve calculated with Eq.~(1).}
\end{figure}

(2) {\it Alpha particles in CdWO$_4$ scintillating bolometer.} 
$QF$ values for $\alpha$ particles
in CdWO$_4$ bolometer were measured in \cite{Arn10} 
using internal trace contaminations in the range of $2.5 - 4.9$ MeV. 
They are shown in Fig.~3a, together with the curve calculated with Eq.~(1) and 
$kB = 8.8$ mg/(MeV$\cdot$cm$^2$). The light yield obtained in a wider energy range
with a smeared $\alpha$ source facing CdWO$_4$ is shown in Fig.~3b (experimental points vs 
present calculations).

\begin{figure}
\centering
\includegraphics[width=8.1cm]{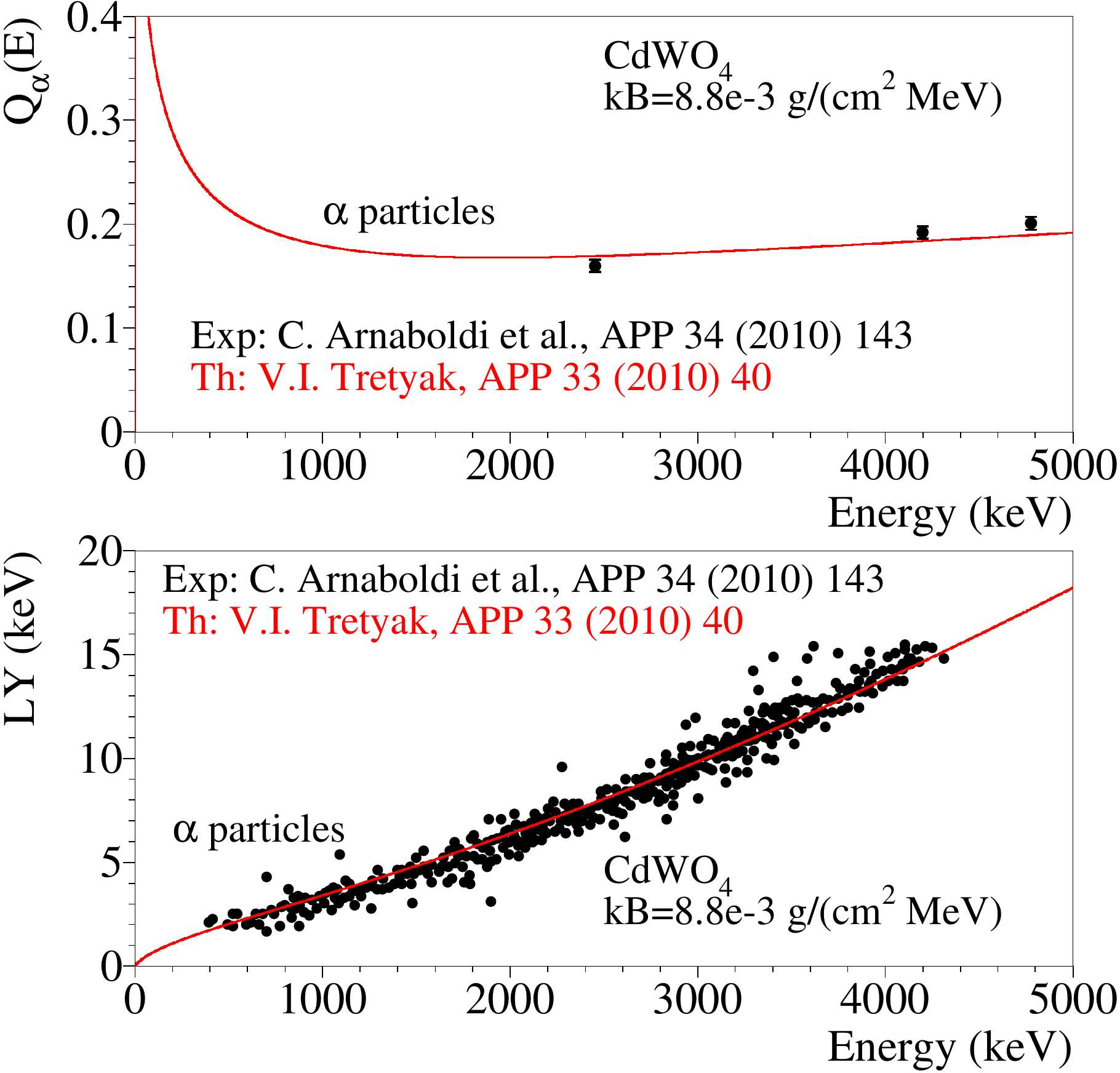}
\caption{Quenching factors for $\alpha$ particles
in CdWO$_4$ scintillating bolometer: (top) experimental $QF$'s \cite{Arn10}
together with fitting curve; (bottom) the light yields measured with a smeared $\alpha$ source:
experimental points \cite{Arn10} and curve calculated with $kB = 8.8$ mg/(MeV$\cdot$cm$^2$).}
\end{figure}

(3) {\it Protons, $\alpha$ particles, Li, C, O, Ti ions in CdWO$_4$ crystal scintillator.} 
$QF$ values for $p$, $\alpha$, and Li, C, O, Ti ions were recently measured with CdWO$_4$
scintillator and ion's beams with $1 - 10$ MeV energies produced by the Tandetron accelerator 
of LABEC at the INFN-Florence \cite{Biz12}. They are shown in Fig.~4.
The data for protons were used to determine the $kB$ value as $kB = 17.4$ mg/(MeV$\cdot$cm$^2$),
and after curves for all other ions were calculated with this $kB$. Fig.~4 shows
general agreement between the experimental results and theoretical dependencies. The biggest
deviation is $\simeq 30\%$ for Ti ions, what could be accepted as satisfactory, especially
for theory with one parameter. It should be also noted that sometimes $QF$'s in WIMP searches are 
known with bigger uncertainties, and thus the present approach could give useful information
on the expected values of $QF$'s.

We would like also to draw attention to difference in quenching factors for $\alpha$ particles
in CdWO$_4$ in Fig.~3 and Fig.~4 (that results also in $kB$ values which are different by factor
of $\simeq 2$). This is one of demonstrations of the statement given in the Introduction:
$QF$'s (and related $kB$) values depend on experimental conditions and cannot be considered as
some fundamental constants for a given scintillating material.

\begin{figure}
\centering
\includegraphics[width=8.1cm]{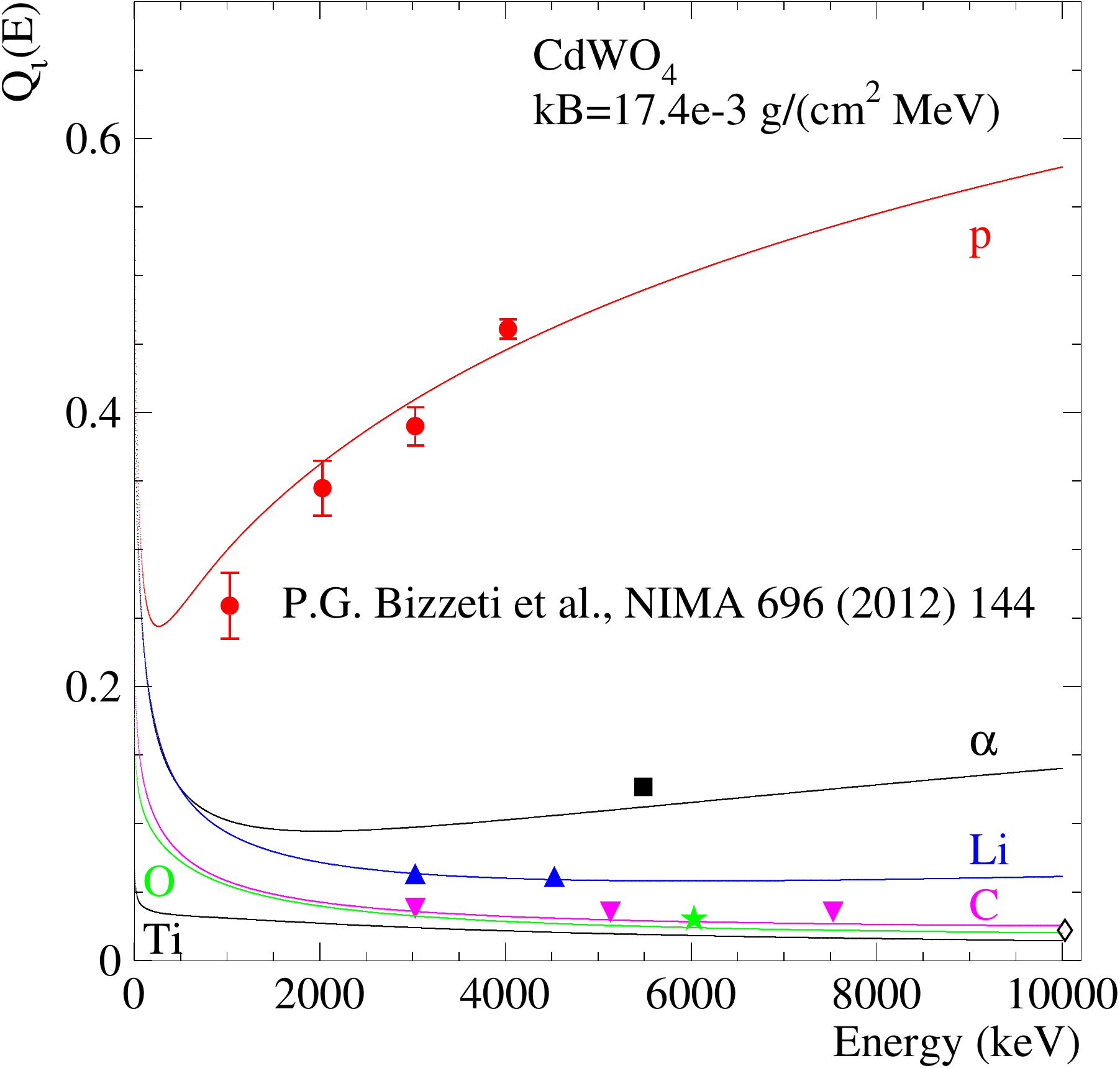}
\caption{Quenching factors for $p$, $\alpha$ particles, Li, C, O and Ti ions
measured with the CdWO$_4$ crystal scintillator \cite{Biz12}. Solid lines represent calculations
with Eq.~(1).}
\end{figure}

(4) {\it Alpha particles in plastic scintillator C$_8$H$_8$.} 
Quenching factors for $\alpha$ particles in solid plastic scintillator C$_8$H$_8$ (BiPo, in R\&D for future
SuperNEMO experiment to search for neutrinoless double beta decay) were measured in \cite{Sar12}. 
They are presented in Fig.~5; calculated curve of $1/QF$ with $kB = 8.5$ mg/(MeV$\cdot$cm$^2$)
is in good agreement with these data.

\begin{figure}
\centering
\includegraphics[width=8.1cm]{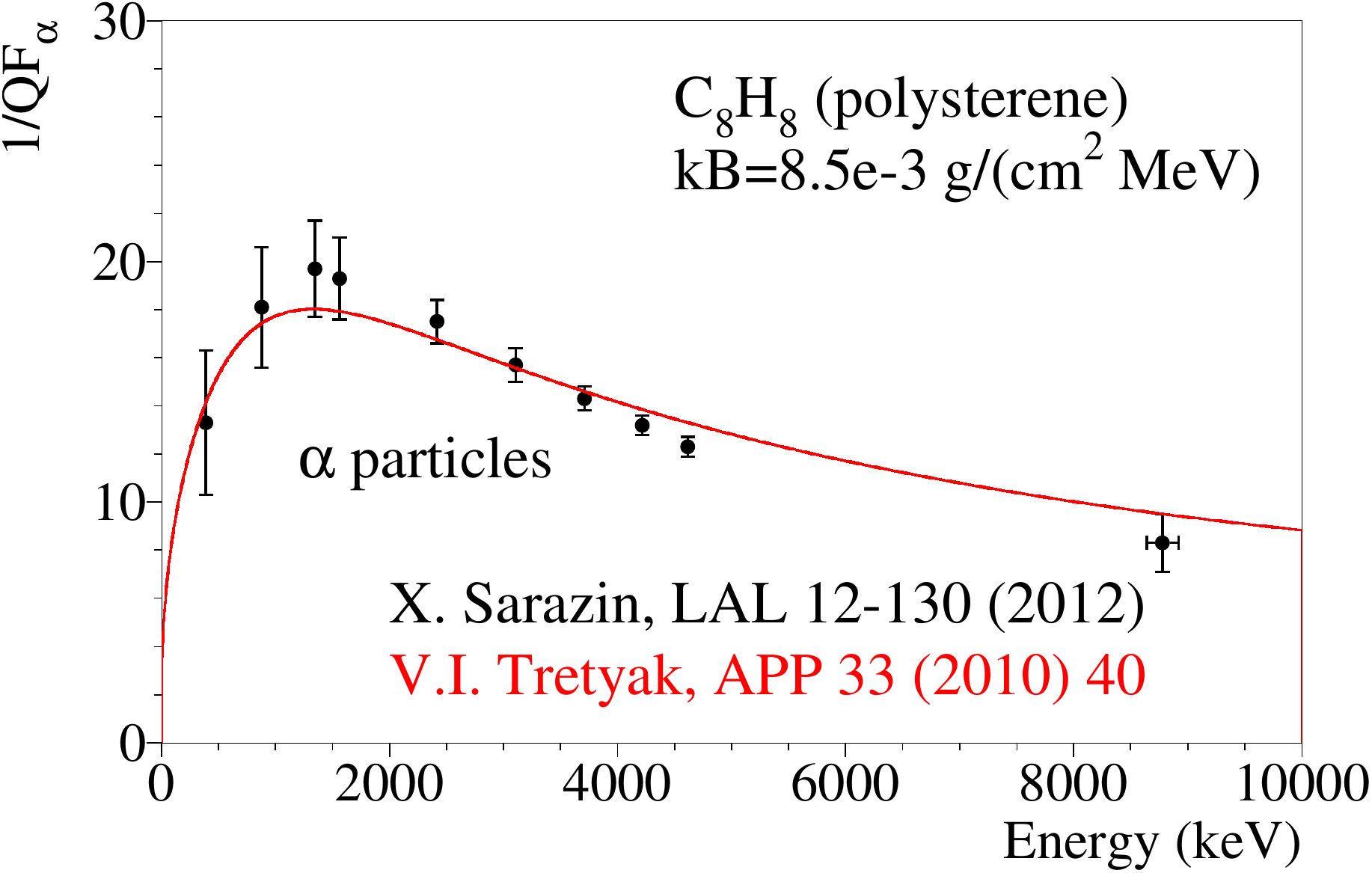}
\caption{Quenching factors for $\alpha$ particles in C$_8$H$_8$
scintillator \cite{Sar12}. Solid line represents calculations with Eq.~(1).}
\end{figure}

(5) {\it Protons in pseudocumene C$_9$H$_{12}$.} 
In our previous work \cite{Tre10}, quenching factors for $\alpha$ particles 
in liquid scintillator C$_9$H$_{12}$ obtained in the BOREXINO experiment \cite{Bac08}
were compared with calculations with Eq.~(1) (see Fig.~3c of Ref.~\cite{Tre10}).
Good agreement was found for the value of $kB = 9.4$ mg/(MeV$\cdot$cm$^2$). 
Later (after publication of \cite{Tre10}) also $QF$'s for protons were measured at
two energies \cite{Bel10}. Corresponding theoretical curve, calculated with the same $kB$ value,
is presented in Fig.~6. As one can see, agreement is excellent, and this gives additional
argument in favour of possibility to describe quenching factors for different particles
with the same $kB$ value if $QF$'s are measured in the same conditions.

\begin{figure}
\centering
\includegraphics[width=8.1cm]{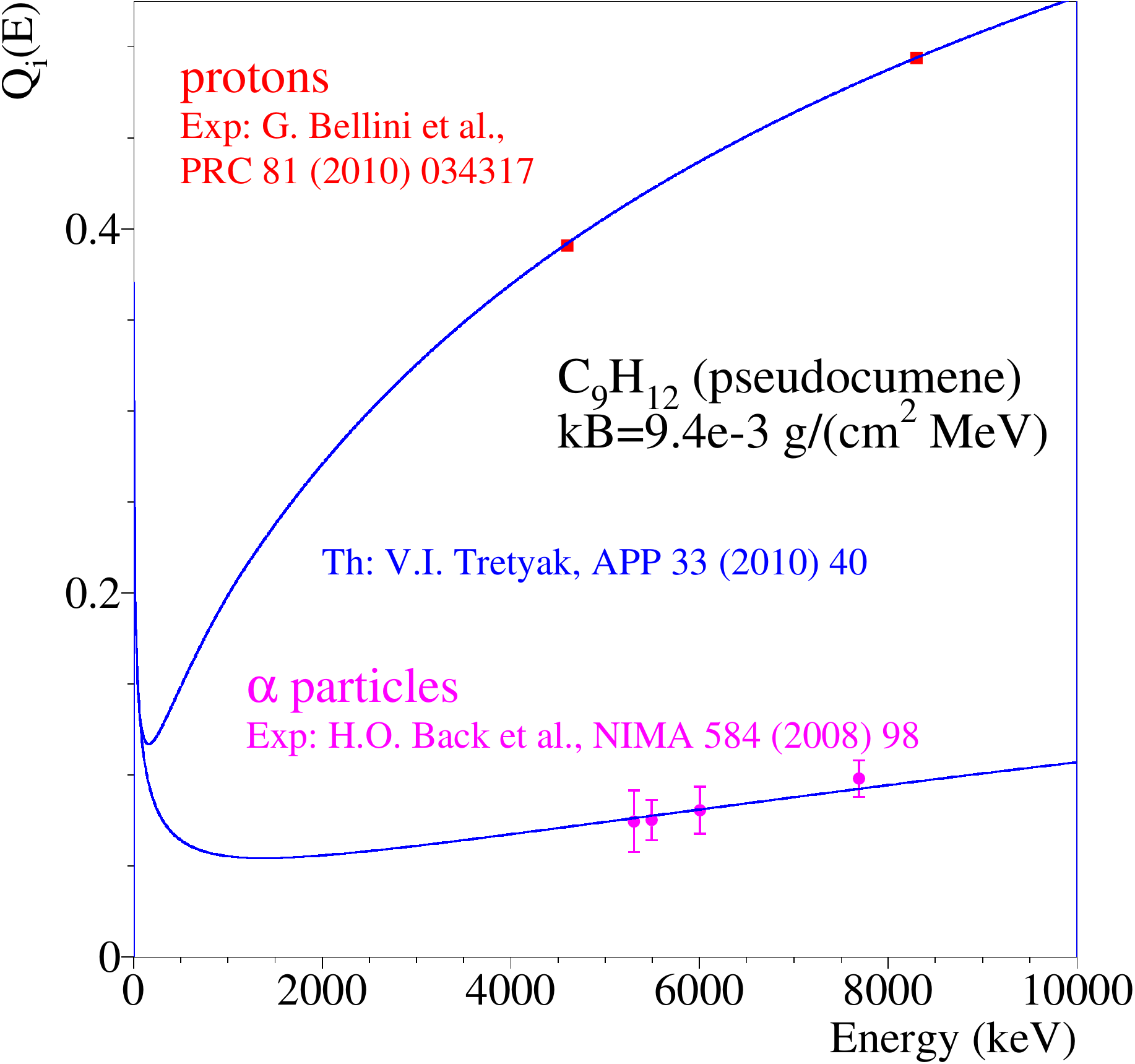}
\caption{Experimental quenching factors for $\alpha$ particles \cite{Bac08}
and protons \cite{Bel10} in pseudocumene C$_9$H$_{12}$ liquid scintillator together
with calculated curves.}
\end{figure}

(6) {\it Quenching factors for Na and I recoils in the DAMA experiment.} 
While we do not present here new calculations of $QF$'s for Na and I recoils in the DAMA 
experiment, we would like to comment once more results obtained already in \cite{Tre10}. 
In the DAMA paper \cite{Ber08}, quenching factors for $\alpha$ particles were derived
using internal trace contamination of the DAMA NaI(Tl) scintillators by U/Th chains.
These $QF_\alpha$'s were obtained in the same conditions as the DAMA dark matter data were
accumulated. Fit of the $QF_\alpha$'s by Eq.~(1) allowed to obtain the value  
$kB = 1.25$ mg/(MeV$\cdot$cm$^2$) and calculate with this value curves for Na and I ions. 
This gave results $QF_{Na} \simeq 0.65$ and $QF_I \simeq 0.35$ at energies $2 - 6$ keV,
much higher than the "standard" values of 
$QF_{Na} = 0.25 - 0.40$ and $QF_I = 0.05 - 0.10$ used in NaI(Tl) DM experiments
(see \cite{Tre10} for corresponding references). However, Figs.~4 and 6 here (together with 
Figs.~3d, 4b, 10a, 10b\&c, 13a in \cite{Tre10}) give arguments that $QF$'s for Na and I recoils 
could be really higher in the DAMA measurements. Higher quenching factors lead to shift of
WIMPs mass in the DAMA measurements to lower values of $\simeq 10$ GeV (see e.g. 
\cite{Bel11,Kel12,Sco13}).

It is also interesting to note that in works \cite{Ari11}, where a combined fit of evidences for 
DM particles in the DAMA, CoGeNT, CRESST, and CDMS-II-Si experiments is described, 
one of the procedures allows $QF_{Na}$ to be a free parameter in the range of $0.2 - 0.6$. 
In this case fit prefers value of $\simeq 0.6$, and possibly even higher value could be preferred
if wider range would be allowed.

(7) {\it Alpha particles in liquid He.} 
Laborious analysis of scintillation yields for electrons and He recoils in LHe was performed recently in
Ref.~\cite{Ito13} for possible use of LHe as a detector in DM experiments. The results are
shown in Fig.~7. We present also a curve obtained in much simpler semi-empirical calculations
with Eq.~(1) which is normalised on single experimental point measured in \cite{Ada01}.
Both the curves are in satisfactory agreement (taking also into account that uncertainties
in calculations of \cite{Ito13} are estimated as 30\% at low energies).

\begin{figure}
\centering
\includegraphics[width=8.1cm]{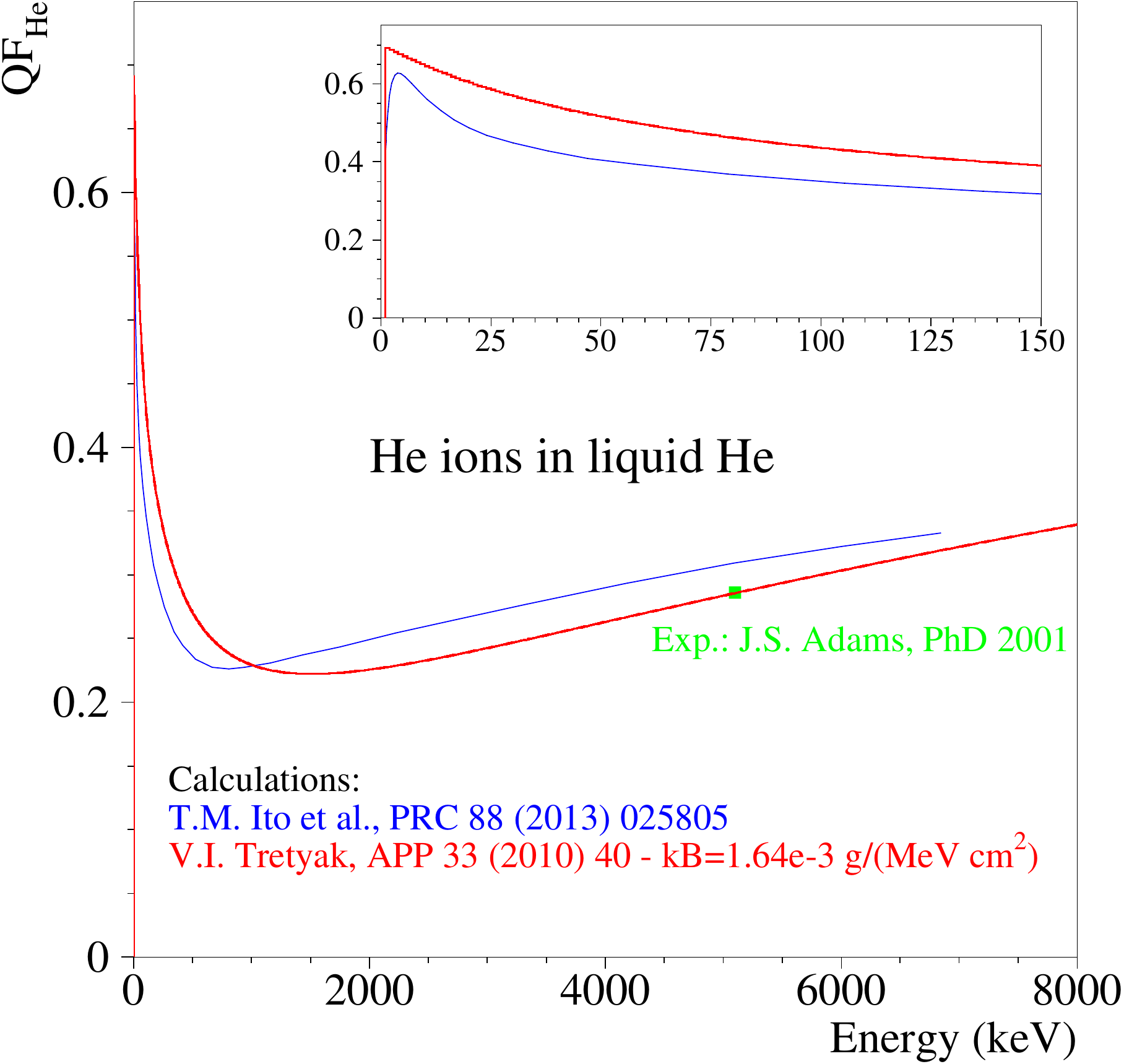}
\caption{Quenching factors for $\alpha$ particles in liquid He 
calculated in \cite{Ito13} (dashed curve) and in the present approach (solid
curve, normalized to experimental point of \cite{Ada01}).}
\end{figure}

\section{Conclusions}

New results of calculation of quenching factors for ions in scintillators in semi-empirical
approach described in \cite{Tre10} are presented here. 
The old Birks formula still gives nice description of $QF$'s for ions in many cases, if the total 
stopping powers for electrons and ions are used, and SP are calculated with the 
ESTAR and SRIM codes which: (a) are publicly available, (b) are ones of the best codes in this field.
There is only one free parameter in the approach: the Birks $kB$ factor. 
It is not considered as some fundamental constant for a given scintillating material 
but as a variable which depends on conditions of measurements and data treatment.

The results presented here give additional arguments in favour of the hypothesis that, once conditions of 
measurements and data treatment are fixed, the $kB$ value is the same for different ions. 
Thus, if $kB$ was determined by fitting data for particles of one kind, 
it can be used to calculate $QF$'s for particles of another kind and for another energies 
of interest. 

Quenching factors for ions calculated in the present approach in general increase at 
low energies, making experimental searches for DM particles more sensitive to low energies
and low WIMPs masses.

\end{document}